\documentclass[aps,manuscript,overcite]{revtex}

\usepackage{graphics}

\input epsf

\tighten
\begin{document}
 


\title{Quantum Fields in a Big Crunch/Big Bang Spacetime}

\author{ Andrew J. Tolley and Neil Turok}

\address{
DAMTP, CMS, Wilberforce Road, Cambridge, CB3 0WA, UK
}

\maketitle

\begin{abstract}
We consider quantum field theory on a spacetime representing
the Big Crunch/Big Bang transition postulated in ekpyrotic
or cyclic
cosmologies. We show via several independent methods that an 
essentially unique matching rule holds connecting the incoming
state, in which a single extra dimension shrinks to zero,
to the outgoing state in which it re-expands at the same rate.
For free fields in our construction there is no particle
production from the incoming adiabatic vacuum. When interactions
are included the particle production for fixed external
momentum is finite at tree level. We discuss a formal
correspondence between our construction and 
quantum field theory
on de Sitter spacetime. 
\end{abstract} 
%

\section{Introduction}

Despite its overwhelming phenomenological
success, the standard big bang cosmology is clearly
incomplete. Its gaps and paradoxes
provide some of the most powerful  
clues to fundamental theory that we possess.
Indeed, it is increasingly evident that 
the real measure of success for  string theory and M theory
will be how well they face up to
the challenges posed by cosmology. Perhaps the 
greatest challenge is that of describing the
initial singularity, a moment 
of infinite density and curvature occurring some 
fifteen billion years ago in our past,
a basic puzzle not
resolved by cosmic inflation.

The 
initial singularity is often
associated with the problem of
the `beginning of time'. But 
the only thing one can legitimately infer from the existence of the singularity
is that
general relativity is incomplete.
Rather than have time
`begin', which is a truly paradoxical notion, or to work with
imaginary time formulations, 
it seems reasonable to explore the alternative possibility
that time may be continued back through the singularity, and even arbitrarily
far into the past.
Such a view is consistent with
what is known so far in string and M theory. Spatial geometry
and topology are only approximate concepts,
as evidenced by orbifold backgrounds\cite{gsw}, and allowed
topology changing processes\cite{flop}.
However, time is built in, in a fundamental role, 
and there is 
no evidence so far that it is allowed to `begin' or `end'.

Recent attempts to construct cosmological scenarios 
employing `brane world' constructions from M theory and string theory  
have led to a re-examination of these issues. 
The `ekpyrotic' scenario\cite{ekpyrotic}, in which a brane collision
is supposed to be the origin of the hot big bang, and its
`cyclic' version\cite{cyclic} in which such collisions occur periodically
into the infinite past and future,
provide alternate approaches 
to the classic cosmological puzzles conventionally
addressed by inflation. In the cyclic model, the
flatness, homogeneity and isotropy of today's Universe is explained
as a consequence of an epoch of vacuum energy domination in the previous
cycle. And the density perturbations needed to seed structure formation
were generated by an inter-brane attractive force 
near the end of the last cycle. An important precursor of these ideas
was the `pre-big bang' model of Veneziano {\it et al.}\cite{pbb}.

The ekpyrotic and cyclic models rest for the most part on
conventional low energy effective field theory and gravity.  
One key event cannot be described
within that approach, namely a collision between
the two end-of-the-world boundary branes (or `orbifold planes').
In the four dimensional effective
description this event appears to be unavoidable,
since the four dimensional effective scale factor is initially
contracting. The four dimensional fields
appearing
in the theory have positive (and growing) kinetic energy and this 
means, through the Friedman equation that 
the contraction cannot be reversed. Within
a finite time one reaches a  `big crunch'
singularity dominated by scalar kinetic energy,
an event which appears at first sight to be irredeemably 
singular.
From the the higher dimensional
viewpoint the situation is  more optimistic.
The geometries of the branes are
regular at the collision and the 
density of matter on the branes
is finite. 
The five
dimensional 
Riemannian curvature is finite everywhere away from 
the singular point. In fact, the only
sense in which the 
higher dimensional geometry is singular is that the
fifth dimension shrinks away to zero size\cite{Seiberg}.

\begin{figure}
{\par\centering 
\resizebox*{4.5in}{3.1in}{\includegraphics{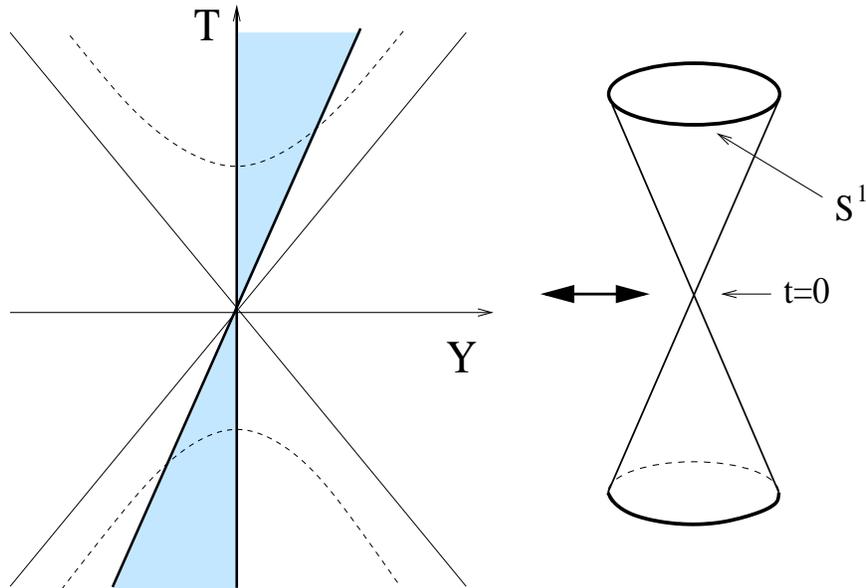}} \par}
\caption{ The compactified Milne universe. On the left is
two dimensional Minkowski space. The Lorentz invariant
coordinate $t$ satisfying $t^2=T^2-Y^2$ is constant on the
dashed surfaces, which provide a spacelike 
foliation of the causal future and past of the origin. 
These surfaces are parameterized by a coordinate $y$.
Identifying $y$ with $y+L$ compactifies space to
produce the spacetime on the right, consisting of two Lorentzian
cones joined tip-to-tip at $t=0$. If the circular sections of
these cones are orbifolded by a $Z_2$, then the two fixed points of
the $Z_2$ are two tensionless branes which collide and pass through
one another at $t=0$.
}
\end{figure}

It is  crucial for the cyclic scenario, as currently formulated, 
that a satisfactory
method be found for passing through the singularity 
corresponding to the collapse of the extra dimension. 
In particular, the issue of matching the density perturbations 
across this singularity has been a matter of fierce debate
\cite{debate}.
A matching rule was proposed in Ref. ~\ref{ekperts}, according to which
the growing mode scale invariant density perturbations developed in the
pre-collapse phase are transmitted across the singularity. But
it is also possible\cite{debate} to match in such a way that only 
the decaying mode is present in the final state. Interesting papers
have subsequently appeared suggesting geometrical methods of
regularizing the singularity\cite{durrer}, or employing 
scalar fields with a negative kinetic term to do so\cite{peters}. However,
none of these methods yet yields a completely unambiguous result
for the case of interest in the ekpyrotic or cyclic scenarios.
We hope that the method developed here 
on more fundamental 
grounds, when extended to include gravitational backreaction, 
will be applicable to
the cosmological case.

Ultimately this issue must be dealt with by string or M theory. Indeed, 
regardless of the ekpyrotic or cyclic scenarios, there 
are good reasons for believing that
this type of singularity {\it must} 
be resolved if string theory is to make sense. 
The shrinking of the extra dimension can be accurately described
using a slow motion (moduli space) approximation, which remains valid 
all the way to zero size. 
The low energy moduli of
string theory and M theory are believed to be fundamental, more
so even than the actions
and Lagrangians they are derived from\cite{moduli}. 
For example, these moduli are
the parameters which interpolate from one corner of M theory to
another. The shrinking away of one extra dimension, in finite time,
seems to be perfectly allowed
in string theory, either if it is one of the nine string theory 
dimensions\cite{Horowitz}, 
or if it is the tenth spatial dimension associated with 
M theory\cite{Seiberg}.
In the former case, the string coupling is constant and may be
taken to be arbitrarily small, so stringy interactions should be
negligible. In the latter case, the string coupling vanishes as
the extra dimension shrinks away. Non-perturbative effects
should, in this case, vanish even more rapidly than perturbative effects.
Thus it is hard to see what would prevent this process. 
The question which must then be faced is: What happens next?

The moduli space description, and the higher dimensional 
picture, both lead to a natural continuation\cite{Seiberg}, 
illustrated in Figure 1. 
The extra dimension contracts to zero at a certain rate
but immediately reappears at the same rate. In the brane picture, 
the two branes collide and pass through one another, a behaviour familiar
from BPS solitons in other contexts. If the collision occurs at finite speed,
one expects some associated 
particle production and consequent back-reaction.

In this paper, we take modest steps towards our eventual goal of a calculation 
of the consequences of a collision between boundary branes in M theory.
There are significant technicalities to be faced even at the 
level of quantum fields, which is all that we shall discuss here.
We shall propose a method of obtaining a
unitary quantum field theory 
on the spacetime illustrated in
Figure 1. Within free field theory, in our construction there is no
particle production in passing from the big crunch
to the big bang phase. However, once interactions
are included, particle production occurs.
For fixed external momenta, the particle production at the big crunch/big bang
transition which
is well defined and finite at tree level. It 
exhibits a power-law fall-off at high momenta
which we argue would likely
be replaced by exponential fall-off in string theory.

\section{Milne and Compactified Milne}

The spacetime we are interested in is a subspace of $d+1$ dimensional
Minkowski space, a trivial solution of $d+1$ dimensional 
general relativity or supergravity. The line element is 
\begin{equation}  
\label{eq:met}
ds^2 = -dT^2+dY^2+d\vec{x}^2,
\end{equation}
where we adopt units in which the speed of light is unity.
We shall refer to $Y$ as the fifth coordinate, having in mind the picture 
that three of the $d-1$
coordinates $\vec{x}$ should provide the 
spatial dimensions of everyday existence, with the remainder 
compactified for example on a torus or orbifold. 
For example in eleven dimensional M theory,
$d=10$ and six of the $\vec{x}$ dimensions
would be taken to be compact.

The line element (\ref{eq:met}) may
be rewritten in terms of new coordinates defined by $T=t\, {\rm cosh} (H_5 y)$,
$Y= t \,{\rm sinh}(H_5 y)$, where $-\infty< t<\infty $ and
 $-\infty< y<\infty $ 
cover 
the causal future ($t>0$) and past ($t<0$) of the origin $Y=T=0$.
We  have here introduced the parameter 
$H_5$, with
dimensions of inverse time. In these coordinates, 
(\ref{eq:met}) becomes
\begin{equation}  \label{metb}
ds^2 = -dt^2 +H_5^2 t^2 dy^2 +d\vec{x}^2, \qquad 
-\infty <t <\infty, \quad -\infty <y <\infty.
\end{equation}
The space comprising the causal future and past of the origin, and 
its light cone $T=\pm Y$
is what we shall define to be the  
Milne universe ${\cal M}$. 
The complement of ${\cal M}$ in Minkowski space comprises the
two Rindler wedges 
$|T|<|Y|$ to the left and right of the origin in Figure 1.

The second step in obtaining the compactified 
Milne universe ${\cal M}_C$ is to compactify the $y$ coordinate into a circle.
Because ${\cal M}$ is invariant under translations, in the quantum theory
there exists a unitary operator $\Lambda(H_5 L)$ implementing 
$y\rightarrow y+L$, which is just
a boost of the original $T,Y$ coordinates on Minkowski space, with 
rapidity $H_5 L$. The coordinate $t$, which is the time in the Milne
universe,  is invariant under this operation.
Let $\Gamma(H_5 L)$ denote the discrete group generated by $\Lambda(H_5 L)$.
Then we define ${\cal M}_C$ to be ${\cal M}/\Gamma(H_5 L)$, i.e. the spacetime
\begin{equation}  
ds^2 = -dt^2 +H_5^2 t^2 dy^2 +d\vec{x}^2, \qquad 
-\infty <t <\infty, \quad 0 <y \leq L,
\label{eq:compa}
\end{equation}
where $y=0$ and $y=L$ are identified. 
We see that 
the parameter $H_5$ is just
the rate of expansion or contraction 
(or `Hubble constant') of the fifth dimension. The space
${\cal M}_C$ is not a manifold, since it is not Hausdorff 
at $t=0$. But of course this is precisely the point of interest
to us. 

So far branes have not entered. We may however further 
reduce the
circle $0<y\leq L$, by identifying its
upper and lower halves under 
the $Z_2$ symmetry $y\rightarrow L-y$. Quantum fields may be
decomposed into 
components which are even or odd under this operation.
The two fixed points of the $Z_2$
symmetry,  $y=0$
and $y=L/2$ can then be viewed as (zero tension) orbifold planes, which
collide and pass through one another at $t=0$ (Figure 1).

We shall also be interested in studying quantum fields in this background 
from the point of view of the dimensionally reduced 
$d$-dimensional theory. 
Writing the $d+1$ dimensional line element as
\begin{equation}  \label{metr}
ds^2 = e^{2\phi \sqrt{(d-2)/(d-1)}} dy^2 + 
e^{-2\phi /\sqrt{(d-2)(d-1)}}g^{(d)}_{\mu \nu} d x^{\mu} d x^{\nu},
\end{equation}
the $d+1$ dimensional Einstein action reduces to that for $d$ dimensional 
gravity with 
a massless, minimally coupled scalar field $\phi$. 
(We adopt units in which the
coefficient of the Ricci scalar in the $d$ dimensional 
Einstein action is ${1\over 2}$.)
The solution ${\cal M}_C \times R^{d-1}$ is now re-interpreted as 
a cosmological solution in which the $d$-dimensional Einstein-frame metric
$g^{(d)}_{\mu \nu}=a^2\, \eta_{\mu \nu}$ with scale factor $a
\propto |t|^{1/(d-2)}$,  and $\phi=
\sqrt{ (d-1)/(d-2)} {\rm ln}|H_5t|$.

It is clear that gravitational waves travelling in the noncompact directions
are minimally coupled both in the $d+1$ dimensional description, 
and in the $d$ dimensional description since the powers of $\phi$ in
(\ref{eq:compa}) were chosen to obtain Einstein-frame 
gravity in the reduced theory. It is straightforward to check that 
a scalar field which is minimally coupled in the $d+1$ dimensional
theory is also minimally coupled in the $d$ dimensional theory.
This means that for the background  
${\cal M}_C \times R^{d-1}$, the dimensionally
reduced action for a minimally coupled scalar $\varphi$ is 
$-{1\over 2} \int \sqrt{-g^{(d)}} g^{(d) \mu \nu} \partial_\mu \varphi
\partial_\nu \varphi =-{1\over 2} 
\int dt |t| \eta^{\mu \nu} \partial_\mu \varphi
\partial_\nu \varphi$, for any $d$.

\section{Free Field Behaviour on ${\cal M}_C$}

Let us now describe 
the  behaviour of free fields on ${\cal M}_C\times R^{d-1}$.
Expanding the fields in plane waves $e^{i\vec{k}\cdot\vec{x}}$ on $R^{d-1}$, 
modes of 
momentum $\vec{k}$ aquire a
mass squared of $\vec{k}^2$ in their two dimensional $(t,y)$ action
or equations of motion. 
The two dimensional line element is just $-dt^2 
+H_5^2 t^2 dy^2$, which is conformally flat, with a conformal factor
which vanishes 
at $t=0$. The two regions $t<0$ and $t>0$
of ${\cal M}$ are each conformal to an infinite cylinder labelled by 
a conformal time $\tau_\pm$, defined by 
$H_5 t= \pm e^{\pm H_5 {\tau_\pm}}$, in the two cases. The line element 
in these coordinates is
then 
\begin{equation}
	ds^2=e^{\pm 2H_5 {\tau_\pm}}(-d\tau^2_\pm+dy^2),
\label{2del}
\end{equation}
where $-\infty < \tau_\pm <\infty$ on each cylinder.
The  conformal factor vanishes as $t$ tends to zero.
In two dimensions the kinetic term for a scalar field is conformally
invariant, and hence does not see the conformal zero. But a
two dimensional 
mass term vanishes like $|t|$.
Therefore, in the  limit $t\rightarrow 0$,
all
field modes behave as those of a massless two dimensional field
on an infinite cylinder. Modes with 
nonzero $y$-momentum $k_y$ oscillate an infinite number of times, as
$e^{\pm ik_y \tau_\pm}$ or $|t|^{ik_y/H_5}$, 
as $\tau_\pm$ tends $\mp\infty$.
On the other hand, modes with $k_y=0$ 
instead evolve linearly in $\tau_\pm$,
which means that they generically diverge as log$|t|$ as
$t \rightarrow 0$. 

The problem of defining a quantum field on ${\cal M}_C$ 
is that of matching the modes across $t=0$, from
their asymptotic behaviours as $t$ tends to zero from above or below.
Since the modes either undergo an
infinite number of oscillations, or are logarithmically divergent, 
this matching is quite subtle. Let us discuss the 
$k_y=0$ modes in more detail.
The general solution for the $k_y=0$ modes behaves as
$\varphi \sim A+B \ln(|H_5 t|)$ as $t$ approaches zero,
with $A$ and $B$ two arbitrary constants.
As we approach $t=0$ the scalar field diverges
logarithmically but its canonically conjugate momentum 
$\pi=|H_5 t| \dot{\varphi}$ tends to
a finite value $H_5 B$. Our problem is then to match a general 
incoming solution
$\varphi_0(t)=A^-+B^- \ln(|H_5 t|)$, $t<0$,  to the corresponding solution for
$t>0$, $\varphi_0(t)=A^++B^+ \ln(|H_5 t|)$. A crude approach would be
to 
simply cut the spacetime off at $t=\pm \delta$, 
identify the field and its conjugate momentum on the two surfaces, and
take the limit of small $\delta$. Since
the momentum is time-independent we obtain
$B^+=-B^-$, independent of $\delta$ as 
$\delta \rightarrow 0$. But matching the field yields
the cutoff-dependent result 
$A^+=A^-+2 B^- \ln(H_5 \delta)$, which 
implies that for any regular in state, the
amplitude of the mode functions generically diverge
logarithmically with the cutoff. If we were to accept this result 
at face value, the
number of particles produced would diverge as the square of the
logarithm of the cutoff. It is tempting to think that this is a
consequence of the unphysical sharp cutoff and that a
smoother regularization prescription might remove the
divergence. However, a smoother cutoff, such as replacing $|t|$ in the
action by $\sqrt{t^2+\delta^2}$ (geometrically, this amounts to
replacing the singular spacetime ${\cal M}_C$ with an hourglass, whose
waist has circumference $\delta H_5 L$),
leads to exactly the same logarithmic divergence as $\delta \rightarrow 0$.
More sophisticated methods must be sought for making quantum field
theory on ${\cal M}_C$ well defined, as we now explain.

\begin{figure}
{\par\centering 
\resizebox*{2.7in}{3.12in}{\includegraphics{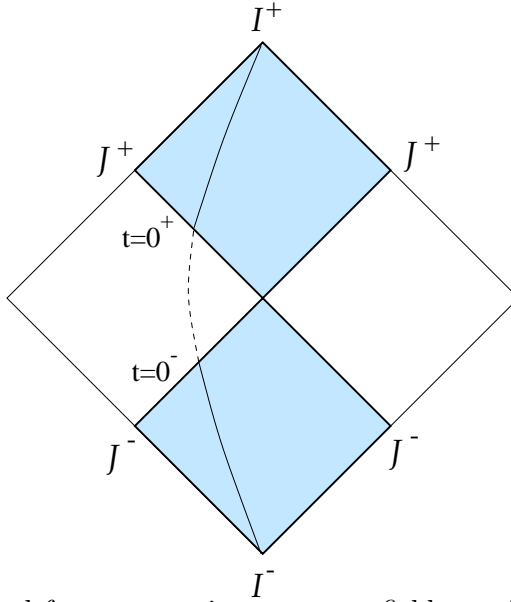}} \par}
\caption{ Our first method for constructing quantum fields on ${\cal M}$,
illustrated in a conformal diagram of Minkowski space. 
The unitary map from the past light cone of the origin, $t=0^-$,
to the future light cone $t=0^+$ is defined by free field evolution 
across the Rindler wedges to the left and right of the origin.
Using this rule we obtain a unitary theory on the Milne universe ${\cal M}$,
which may then be compactified into the space ${\cal M}_C$ shown in Figure 1.
}
\end{figure}

\section{Quantum Field Theory on ${\cal M}_C$}

We shall describe several different
constructions for quantum 
fields on ${\cal M}_C$, which all yield an 
essentially unique result. 

The first method is based on Figure 2. We use the embedding of
${\cal M}_C$ in Minkowski spacetime to define the map from $t=0^-$
to $t=0^+$. This is possible as 
long as one or more of the
$\vec{x}$ directions are noncompact, because in this case,
the corresponding momenta
$\vec{k}$ are continuous and $\vec{k}=0$ is a set of measure
zero. 
From the two dimensional standpoint, this means that all
modes are effectively massive. And for massive fields, free field
evolution provides a unitary map between the past light cone of
the origin ($t=0^-$) and the future light cone ($t=0^+$), because
no information can be carried off to null infinity $J^+$. 
The coordinate $t$ analytically continues to a spacelike
coordinate in the Rindler wedges, and as one follows
the trajectory plotted in Figure 2, this coordinate 
runs from zero to a finite value then back to zero.
So in effect a `clock' measuring $t$ makes no progress 
whilst the trajectory is outside the region ${\cal M}$ of interest. 

\begin{figure}
{\par\centering
\resizebox*{4.5in}{2.1in}{\includegraphics{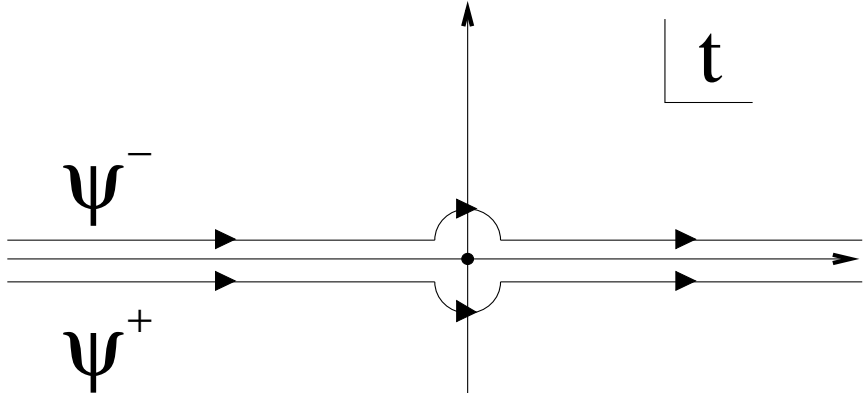}} \par}
\caption{ Integration contours used to define the positive and negative 
frequency modes on the entire Milne universe ${\cal M}$. 
}
\end{figure}

This first method may also be viewed as a certain
analytic continuation in the complex
$t$-plane (Figure 3).
The field equation is analytic in the original Minkowski
coordinates $T$ and $Y$ so the global solution may be obtained unambigously
by analytic continuation in those coordinates. We shall show that
this corresponds to the
continuation of the
positive and negative frequency mode functions $\Psi^{(+)}$ and
$\Psi^{(-)}$ from
negative to positive $t$, 
illustrated in the
diagram above. The positive frequency mode functions so defined
are analytic in the lower half $t$-plane, and the negative frequency
mode functions are analytic in the upper half $t$-plane.
The quantum field, being a sum of the positive and negative frequency modes,
is continued in this
mixed fashion across $t=0$. This analytic continuation method
is actually more fundamental than continuation across the Rindler wedges,
because it does not involve those unphysical regions. This is an
important distinction when we introduce interactions. There
is an ambiguity (for example about what the mass used in the
free field propagation should be) in the Minkowski space
continuation, but no corresponding ambiguity in the 
method illustrated in Figure 3.

Nevertheless it is interesting to discuss in more
detail how the two methods correspond, for free fields.
The coordinate $t$ continues to a spacelike variable
$t=\pm is$ in the Rindler regions, where the line element is $ds^2 -s^2 dy^2$, 
and $y$ is now timelike. So in the Rindler regions, the
continuation across from $t=0^-$ to $t=0^+$ occurs via paths which
run up (or down) the imaginary $t$ axis and back again. On these paths,
$y$ is also evolving from $-\infty$ to $+\infty$. Modes with
nonzero momentum $k_y$ in the Milne region undergo an infinite number
of oscillations as they approach 
$t=0^-$ from above or below, and an infinite number more as they cross 
the Rindler wedges. More subtle is the behaviour of the
$k_y=0$ modes. As we discussed, these modes generically diverge logarithmically 
as one approaches $t=0^-$. By a choice
of phase one can put this divergence into 
the imaginary part of the mode functions.
Then, as one circumnavigates the origin in the complex $t$-plane, the
logarithm aquires an imaginary part of $\pm i\pi$. This causes 
the real part
of the mode functions to undergo a jump, of
just the amount needed to
reverse its sign. This is illustrated in Figure 4. 

\begin{figure}
{\par\centering
\resizebox*{4.5in}{3.5in}{\includegraphics{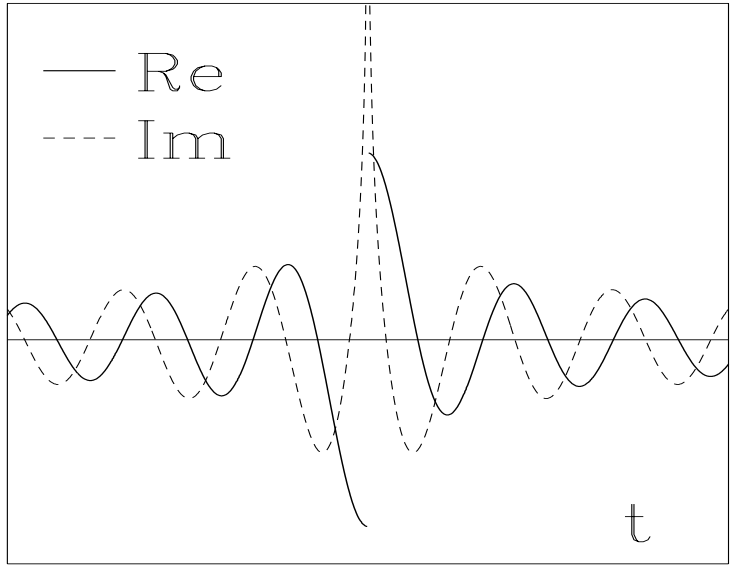}} \par}
\caption{ Behaviour of the $k_y=0$ modes as they cross $t=0$. 
With our choice of phase, the imaginary part (Im) diverges logarithmically
but the real part (Re) is finite. Analytic continuation along the path
shown in Figure 3 causes the real part to be
odd in $t$ whereas the
imaginary part is even.
}
\end{figure}

The method described above is, we believe, completely adequate for dealing
with quantum fields on ${\cal M}_C$. However, it is also interesting
an important to develop the corresponding description of
passage through the singularity in the $d$ dimensional 
effective theory. In this theory, 
a scalar field has action
\begin{equation}
- \int dt d^{d-1} \vec{x} |t| {1\over 2} \eta^{\mu \nu} \partial_\mu \varphi
\partial_\nu \varphi,
\end{equation} 
with a specific time dependence in the kinetic term. Our approach here
will be to regularize the theory by changing $|t|$ to
$|t|^{1-\epsilon}$,
with $\epsilon$ a parameter analogous to that in dimensional regularization,
to be taken to
zero after renormalization. It is then necessary to 
add counterterms to the Hamiltonian at $t=0$ in order to 
render the time evolution operator well defined in the $\epsilon \rightarrow 0$
limit.
These counterterms have the effect of inducing a shift in the scalar field,
proportional to its momentum, and 
analogous to the jump produced in the analytic continuation method
illustrated in Figure 4. This shift cancels divergences and renders the final
state well defined. 
We shall show that within this method, 
demanding that the counterterms be local in $\vec{x}$, and 
imposing time reversal symmetry is enough
to uniquely fix the vacuum state. 

Finally, we shall point out an intriguing mapping between this problem and
that of free fields on de Sitter spacetime. Under this mapping,
the surface 
$t=0^+$ corresponds to the past timelike infinity in de Sitter space,
and $t=0^-$ corresponds to future timelike infinity. While these
two surfaces are only connected at a point in the Milne
universe, they are connected by a smooth bulk (comprising the 
entire spacetime)
in the de Sitter case. The matching in de Sitter spacetime is 
unambiguous and again we shall show it corresponds to the previously 
obtained results. There are holographic elements of this correspondence.
Holography is naturally framed in terms of null surfaces\cite{bousso}, and
our approach involves 
matching information located on
the two null surfaces $t=0^-$ and $t=0^+$. However, when we map to 
de Sitter spacetime, these two surfaces map to two spacelike surfaces, 
future
and past timelike infinity, which are those which have been
employed in 
the proposed
de Sitter-CFT correspondence \cite{strominger}.

All of these methods yield the same result for the quantum vacuum state on
${\cal M}_C$. Because there is no mixing 
of the positive and negative frequency
modes,
there is no particle production in the free field theory.
However, once interactions are included, particle production
occurs, and in Section VI we demonstrate that it is well
defined. The $k_y\neq 0$ modes are produced with a density
which tends to zero exponentially as $H_5$ vanishes, 
suggesting an adiabatic limit in which the particle
production vanishes in the limit of slowly colliding boundary
branes. The $k_y=0$ modes do not show this behaviour,
but we shall discuss how within string theory we can anticipate 
how an adiabatic limit may in fact emerge.

Finally, let us mention the connection between this work and other,
more ambitious attempts to directly construct  
string theory on the compactified Milne spacetime
considered here. 
Nekrasov\cite{Nekrasov} considered string theory on the Lorentzian orbifold
constructed by orbifolding Minkowski
spacetime by a boost.  
In that construction, the two Rindler wedges become compactified in 
a timelike direction, producing two extra cones projecting
horizontally from the origin, which 
possess closed timelike curves.
Additionally, line segments emanating from the 
origin are produced in each of the four null directions. 
Cornalba and Costa
avoid these features by modding out by a boost combined with a translation,
replacing them instead by a new region containing a 
naked timelike singularity\cite{cornalba}. Balasubramanian {\it et al.} 
consider other examples of time-dependent orbifold backgrounds in string 
theory. Whilst free strings seem to be well defined in these backgrounds, 
it is not yet clear whether interactions can be consistently
introduced. 

The approach we suggest here does not amount to orbifolding 
Minkowski space. Instead, we use free field evolution 
(or, equivalently analytic continuation in $t$) to define a matching
rule between the big crunch and the big bang. This difference is
unimportant in the free theory, since the only difference in that
case between our approach and the orbifold approach is that we
would declare that the extra regions in the orbifold approaches 
do not exist. It is when we introduce interactions that the
difference becomes crucial. In our case, the interaction 
Lagrangian is only integrated over the physical compactified
Milne spacetime, whereas in the orbifold approaches it would
be integrated over the additional regions too, containing
closed time-like curves or naked singularities. We should
stress that we have not attempted to construct string theory
in our approach, therefore we cannot say whether string theory will
ultimately be consistent on compactified Milne. 
However the field theory results are
suggestive and we hope they will be a guide to such a construction.

\section{Embedding Milne in Minkowski}

\subsection{Positive and Negative Frequency modes}

In this section we describe our first construction of quantum field theory
on ${\cal M}_C\times R^{d-1}$. A Fourier mode of a massless field,
$\varphi(t,y,\vec{x}) = \varphi(t)
 e^{i(\vec{k}\cdot \vec{x}+k_y y)}$, obeys the field equation
\begin{equation}
\ddot{\varphi} + {1\over t} \dot{\varphi} +{k_y^2\over H_5^2 t^2} \varphi
+m^2 \varphi=0,
\label{eq:mode1}
\end{equation}
where  dot
denotes partial derivative with respect to
$t$. We have introduced the effective two dimensional 
mass $m^2\equiv \vec{k}^2$, and henceforth the $\vec{x}$ dependence shall
play a purely spectator role. 
Equation (\ref{eq:mode1})
is just Bessel's equation, with imaginary order $\nu=
i k_y/H_5$. It has a singular point at $t=0$.  The solutions which
tend to positive and negative frequency WKB modes at late times
are the Hankel functions, and the properly normalised 
outgoing positive and negative 
frequency modes are 
\begin{equation}
\psi^+_{k_y}=\sqrt{\frac{\pi}{4H_5}} 
e^{{\pi k_y\over 2H_5}} H^{(2)}_{ik_y/H_5}(mt)e^{i k_y y}, \qquad
\psi^-_{k_y}=\sqrt{\frac{\pi}{4H_5}}
e^{-{\pi k_y\over 2H_5}} H^{(1)}_{ik_y/H_5}(mt)e^{i k_y y}.
\label{eq:posneg}
\end{equation}
We would like to continue these modes to negative times. 
The Hankel functions have the following integral representations
\begin{equation}
	H^{(1)}_{\nu}(z)= 
	\frac{e^{-{i \pi \nu \over 2}}}{\pi i}
	\int_{-\infty}^{\infty} du \exp({iz \cosh u -\nu u}),
\label{eq:h1}
\end{equation}
which is analytic in the upper half $z$-plane, and 
\begin{equation}
	H^{(2)}_{\nu}(z)=-\frac{e^{{i \pi \nu \over 2}}}{\pi i}
	\int_{-\infty}^{\infty} du \exp({-iz \cosh u -\nu u}),
\label{eq:h2}
\end{equation}
which is analytic in the lower half $z$-plane.
Consequently the Milne mode functions can be 
expressed as
\begin{equation}
	\psi^+_{k_y}=\frac{i}{2} \sqrt{\frac{1}{\pi H_5}}
	\int_{-\infty}^{\infty} du \exp({-imt \cosh u -i\frac{k_y}{H_5} (u-H_5y)}), \qquad \psi^-_{k_y}=\psi^{+*}_{k_y}.
\end{equation}
By shifting the integration variable $u \rightarrow u+yH_5$, we obtain
\begin{equation}
	\psi^+_{k_y}=\frac{i}{2} \sqrt{\frac{1}{\pi H_5}}
	\int_{-\infty}^{\infty} du \exp({-imt \cosh u \cosh H_5y -imt \sinh u
	\sinh H_5y -i\frac{k_y}{H_5} u}).
\end{equation}
Further changing variables to $K_Y=-m \sinh(u)$ gives
\begin{equation}
	\psi^+_{k_y}=\frac{i}{2} \sqrt{\frac{1}{\pi H_5}}
	\int_{-\infty}^{\infty} \frac{dK_Y}{\sqrt{K_Y^2+m^2}} 
e^{i {k_y\over H_5}
	\sinh^{-1}(K_Y/m)} \exp(iK_Y Y -i \sqrt{K_Y^2+m^2} T),
\label{eq:posfreq}
\end{equation}
where $T=t \cosh(H_5 y)$ and $Y =t \sinh(H_5y)$ are the embedding
co-ordinates in Minkowski space. This is a superposition of positive
frequency plane wave modes on Minkowski space with momentum
$K_Y$. We note that the right hand side is an oscillatory integral 
which can be defined for $t<0$ by inserting a suitable convergence factor.
Therefore the integral representation 
 may be used as the definition 
of the mode functions there.

The above integral representations 
of the Hankel functions define a natural 
analytic continuation across $t=0$. One can read off from 
(\ref{eq:h1}) and (\ref{eq:h2}) the relations 
\begin{eqnarray}
	H^{(2)}_{\nu}(e^{-i\pi} z)=- e^{i\pi \nu} H^{(1)}_{\nu}(z), \qquad 
	H^{(1)}_{\nu}(e^{i\pi} z)=- e^{-i\pi \nu} H^{(2)}_{\nu}(z).
\label{eq:rule1}
\end{eqnarray}
To see what these imply for the $k_y=0$ modes, 
recall that 
\begin{equation}
        H^{(1)}_0(mt) \equiv J_0(t) +i N_0(t),
\qquad H^{(2)}_0(mt) \equiv J_0(t) -i N_0(t).
\label{eq:hank}
\end{equation}
The rule (\ref{eq:rule1}) implies that the analytic continuation
of $H^{(2)}_0(mt)$ to negative values is $-J_0(-t) -i N_0(-t)$.
Therefore from (\ref{eq:hank}) 
the real part of $H^{(2)}_0(mt)$ is an odd function of 
$t$, with a discontinuity at $t=0$, and the imaginary part is
even, with a logarithmic divergence at $t=0$. The real and imaginary parts
are illustrated in Figure 4.

From the integral representations (\ref{eq:h1}) and (\ref{eq:h2}) one can
determine the behaviour of the analytically continued Hankel functions at 
large positive or negative $t$ by performing the integral via the
stationary phase method, obtaining
\begin{equation}
  H^{(2)}_{ik_y/H_5}(mt) \sim - {e^{-\pi k_y\over 2}\over i\pi} 
 e^{-imt} \sqrt{2 \pi\over|mt|} e^{\mp i\pi\over 4} \quad t\rightarrow \pm \infty, 
\end{equation}
and similarly for $H^{(1)}_{ik_y/H_5}(mt)$, with $i\rightarrow -i$ and
$k_y\rightarrow -k_y$.
This continuation implies that there is no particle production
since positive frequency incoming modes are matched to
only positive frequency outgoing modes.

It is important to stress that this choice of vacuum is priviledged. As
we explained earlier if we cut off the singularity in a crude way,
then for a generic
choice of $|in>$ and $|out>$ states, for each $\vec{k}$ 
we would obtain in the
$k_y=0$ mode a particle production rate that diverges logarithmically
with the regulator. It is only for the special case in which we define
$|out>=|in>$ or at least some finite Bogoliubov transformation of the
$|in>$ state that we obtain a finite result. From the $d$ dimensional
perspective this seems contrived, but from the $d+1$ dimensional
picture and the embedding in Minkowski space it is clearly
the most natural choice. In later sections we shall also 
justify this matching from a purely $d$ dimensional point of view. 

Finally, let us mention that our 
definition of in and out vacuum modes is {\it not} the same as
that which has conventionally been used in treatments of quantum fields on
Milne spacetime (see for example, Ref.~\ref{chitre}). In previous work,
only the $t>0$ part of ${\cal M}$ was used, and the initial vacuum was taken
to be the
`conformal vacuum' as $t \rightarrow 0^+$, defined by 
the `positive frequency' modes behaving as 
$e^{-i k_y \tau}$ as the conformal time 
$\tau \rightarrow
-\infty$. This is of course, not an adiabatic vacuum state, and therefore
a somewhat arbitrary choice. 
In the conformal vacuum state, one finds particle production occurs in passing
from the big bang $t\rightarrow 0^+$ to the asymptotic future,
even in free field theory\cite{chitre}.

\subsection{Projection onto ${\cal M}_C$}

We have not yet distinguished between the Milne space ${\cal M}$ and its
compactification ${\cal M}_C$, which as we
 described above equation (\ref{eq:compa}) is just
${\cal M}/\Gamma(H_5 L)$, with  $\Gamma(H_5 L)$ the group of boosts 
with rapidity $H_5 L$.

%

In Minkowski spacetime a particle is defined in a group
theoretic sense as an irreducible projective representation of the
Poincare group. 
We can similarly define particles on ${\cal M}_C$ 
by using representations on the covering Minkowski space 
that are invariant under
the action of the boost $\Gamma(H_5 L)$. The map from Milne to Minkowski
introduced in the previous section is inverted by means of a Fourier transform,
to obtain
\begin{equation}
	\Psi_{K_Y}(Y,T)=e^{iK_Y Y-i \sqrt{K_Y^2+m^2} T}
=\int_{-\infty}^{\infty} \frac{dk_y}{2 \pi} e^{\pi k_y\over 2 H_5} 
H^{(2)}_{i k_y/H_5}(mt)
	e^{ik_y (y-H_5^{-1} \sinh^{-1}(\frac{K_Y}{m}))}
\end{equation}
The plane waves $\Psi_{K_Y}(Y,T)$ form a representation of the two dimensional
Poincare group, and the action of the boost $\Lambda$ 
on these modes can be
expressed as
\begin{equation}
	\Lambda : \Psi_{K_Y}(Y,T) \rightarrow \Psi_{K_Y}(\Lambda
	Y,\Lambda T)
\end{equation}
where $(\Lambda Y,\Lambda T)=(t \sinh(H_5(y+L)), t \cosh(H_5(y+L)))$
is simply a translation by $L$ in $y$. A representation of the group
$Poincare/\Gamma(H_5 L)$ can be constructed by simply summing over all
boosts,
\begin{eqnarray}
	\tilde{\Psi}_{K_Y}(Y,T)=\sum_{n=-\infty}^{\infty}
 \Psi_{K_Y}(\Lambda^n(H_5L) Y,\Lambda^n(H_5L) T)=
	\sum_{n=-\infty}^{\infty}
	 \Psi_{K_Y}(\Lambda(nH_5L) Y,\Lambda(nH_5L) T).
\end{eqnarray}
We shall only use these functions on the physical region of interest,
namely ${\cal M}_C$, where they are given by
\begin{equation}
	\tilde{\Psi}_{K_Y}(Y,T)=\sum_{n=-\infty}^{\infty}
	\int_{-\infty}^{\infty} \frac{dk_y}{2 \pi} 
e^{\pi k_y\over 2 H_5} 
H^{(2)}_{i k_y/H_5}(mt)
e^{ik_y (y+nL-H_5^{-1} \sinh^{-1}(\frac{K_Y}{m}))}.
\end{equation}
Now using the Poisson summation formula 
$\sum_{n=-\infty}^{\infty} \int_{-\infty}^{\infty}  dt
	e^{2\pi int} f(t) =
	\sum_{m=-\infty}^{\infty} f(m)$,
we obtain
\begin{equation}
	\tilde{\Psi}_{K_Y}(Y,T)=\frac{1}{L} \sum_{n=-\infty}^{\infty}
	e^{\pi^2 n\over H_5 L} H^{(2)}_{ 2 i\pi n / H_5 L}(mt)
	e^{i{2 \pi n\over L} (y-H_5^{-1} \sinh^{-1}(\frac{K_Y}{m}))}.
\end{equation}
This is just the expected result that summing over boosts projects
out only those states that are translation invariant under $y
\rightarrow y+L$, and is equivalent to quantizing the momentum $k_y
=2\pi n/L$. If we were to perform the further projection onto the
orbifold $S_1/Z_2$ mentioned in the introduction, we would now consider
separating the $k_y$ modes into those which are 
odd and even under $y\rightarrow
L-y$. In string theory, this step introduces new states (`twisted states')
but for field theory describing quantum mechanical particles, 
it has no such effect.

The Feynman propagator on ${\cal M}$ is obtained by simply
restricting the $d+1$ dimensional Minkowski space propagator 
to the Milne region. In $d+1$ dimensions the
Feynman propagator is \cite{Birrell}
\begin{equation}
	G_F(x,x')=-i{\pi \over (4 \pi i)^{d+1\over 2} } 
\left({m^2 \over 
\sigma +i\epsilon}\right)^{d-1\over 4}  
	H_{(d-1)/2}^{(2)}( m(-\sigma-i\epsilon)^{1/2}),
\end{equation}
where $x=(T,Y,\vec{x})$ and 
$\sigma =(\vec{x}-\vec{x}'^2)+ (Y-Y')^2-(T-T')^2$. 
Restricting the Feynman propagator to the Milne patch simply requires
writing $\sigma$ in terms of the Milne co-ordinates
$\sigma(x,x')=(\vec{x}-\vec{x}')^2-(t-t')^2-4 t t' \sinh^2(H_5(y-y')/2)$. 
The Feynman propagator on the compactified Milne spacetime is obtained
by projecting onto the boost invariant states. 
This is given by
\begin{eqnarray}
	G^{{\cal M}_C}_F(x,x')&=&
\sum_{n=-\infty}^{\infty} G^{\cal M}_F(\Lambda^n(H_5 L)x,x')\cr
&=&\sum_{n=-\infty}^{\infty} -i{\pi \over (4 \pi i)^{d+1\over 2} } 
\left({m^2 \over 
\sigma_n +i\epsilon}\right)^{d-1\over 4}  
	H_{(d-1)/2}^{(2)}( m(-\sigma_n-i\epsilon)^{1/2})
\end{eqnarray}
where $\sigma_n(x,x')=(\vec{x}-\vec{x}')^2-(t-t')^2-4 t t' \sinh^2(H_5(y+nL-y')/2)$. The Feynman propagator, in addition to the interaction vertices,
is all one needs in order compute the $S$ matrix via perturbation theory
on ${\cal M}_C$. 


\subsection{UV divergence behaviour}

It
is important to understand whether compactifying Minkowski spacetime
into ${\cal M}_C\times R^{d-1}$ introduces any new ultraviolet divergences.
%
For the construction given above, the free field propagator on
${\cal M}$ is just the 
Minkowski space propagator evaluated on ${\cal M}$. Therefore
it has just the usual divergences. 
In this section we shall show that the same is true for the propagator
on ${\cal M}_C$, 
for all points $x$ and $x'$ away from $t=0$. This is to be
expected intuitively since 
the Green
functions on ${\cal M}_C$ are constructed by 
summing over boosts on one argument 
$x$, and these boosts carry $x$ further and further from 
$x'$. 

The difference
between the Feynman propagator on $\cal{M}_C$ and $\cal{M}$ 
is given in $d+1$ dimensions by
\begin{equation}
	\Delta G_F(x,x')=\sum_{n=-\infty, \neq 0}^{\infty}  
-i{\pi \over (4 \pi i)^{d+1\over 2} }\left({m^2 \over
\sigma_n +i\epsilon}\right)^{d-1\over 4}
        H_{(d-1)/2}^{(2)}( m(-\sigma_n-i\epsilon)^{1/2}),
\end{equation}
which in the coincidence limit $x'=x$ becomes
\begin{equation}
2 \sum_{n=1}^{\infty}
-i{\pi \over (4 \pi i)^{d+1\over 2}} \left({m^2 \over
-4 t^2 \sinh^2(H_5nL/2) +i\epsilon}\right)^{d-1\over 4}
        H_{(d-1)/2}^{(2)}(m ( 4 t^2 \sinh^2(H_5nL/2)-i \epsilon)^{1/2}).
\end{equation}
The large $|z|$ asymptotic behaviour of the Hankel function is 
$H_{\nu}^{(2)}(z) \rightarrow 
(\frac{2}{\pi z})^{1/2} e^{-i(z-\frac{1}{2} \nu
\pi-\frac{1}{4} \pi)} $ and so the sum is rapidly
convergent for nonzero $t$. Thus, at least away from $t=0$ the 
UV divergence behaviour of the Green function on ${\cal M}_C$ 
is just the same as that on Minkowski space. The behaviour 
of the Green function at $t=0$ is a more delicate matter,
linked to the way in which interactions enter, which we shall
discuss below. 

In the next section we shall see that if interactions are introduced
on ${\cal M}$ as integrals over fields on ${\cal M}$, there are physical
processes such as particle creation from the vacuum that occur at
tree level, and which have no counterpart in Minkowski spacetime. They arise
because energy is no longer conserved when the interactions are
time-dependent.

\section{Interacting Field Theory}

The prescription discussed above for matching the big crunch phase
to the big bang phase in the Milne universe relied on
free field theory. That is, in the Minkowski spacetime within
which the Milne universe is embedded, we are propagating the
fields according to the free field equations from the past light cone
(on which Milne time is $t=0^-$)
to the future light cone (on which Milne time is $t=0^+$). With
this prescription, as we have emphasized, there is no
particle production. However, once interactions are included,
particles are generically produced because the interaction terms
in the Hamiltonian are time dependent. We shall calculate this effect 
in this section, using a very
simple toy model of the interactions. This is not 
intended to accurately represent the actual interactions in string
theory, but we hope will illustrate the general behaviour including the
sensitivity to infrared and ultraviolet cutoffs. The former
should come from cosmological evolution, since the growth of
the extra dimension ceases when the universe becomes radiation dominated.
Ultraviolet divergences have to be controlled by string theory
or M theory effects and we shall comment on the possible form of these below. 
It is important to stress that we use the Minkowski embedding only
to determine a matching condition in the free field theory. The interacting
theory lives on the physical spacetime ${\cal M}_C$. This is the
sense in which our approach differs from one employing an interacting 
theory on a Lorentzian orbifold 
which is Minkowski space modulo a boost.

As a very simple example, consider an interaction of the form
${\cal S} = - \int {1\over 2} \mu^2 \varphi^2$, where the integral runs
only over ${\cal M}_C$. For concreteness we shall take $d=4$, so there
are three noncompact dimensions $\vec{x}$. 
The interaction is simply a mass term, which from
the point of view of the embedding theory in Minkowski space,
is turned off outside the future and past light cones. We would
like to compute the particle production due to this interaction.
The quantum field $\varphi$ is expanded in terms of creation and
annihilation operators as 
\begin{equation}
\varphi= \int {d^3\vec{k} \over (2\pi)^3}\sum_{k_y} \left[a_{{k_y},\vec{k}}
\psi^+_{k_y} 
e^{i{k_y}y +i \vec{k}.\vec{x}} 
+{\rm h.c.}\right],
\label{phic}
\end{equation}
where the $\vec{k}$ dependence of the modes is not explicitly shown.
The creation and annihilation operators are normalized to obey
\begin{equation}
\left[a_{{k_y},\vec{k}}, a_{{k'_y},\vec{k'}}^\dagger\right]=
 \delta_{{k_y},{k_y'}} (2\pi)^3 {\delta}^3(\vec{k}-\vec{k'}).
\end{equation}
We may now compute the transition amplitude between the incoming 
vacuum state and an outgoing state with two particles, with
equal and opposite momenta $k_y$ and $\vec{k}$. The calculation is
straightforwardly performed by first integrating over $y$ to obtain the
delta function $\delta_{k_y+k_y',0}$ corresponding to momentum 
conservation. Then we use the representation of the Hankel functions
given in (\ref{eq:posfreq}), evaluated at $Y=0$, $T=t$, to obtain the
interaction matrix element
\begin{equation}
{-i\mu^2 \over 8 H_5 \pi V} (2 \pi)^3 \delta^3(\vec{k}+\vec{k}') 
\int dt |t| \int\int du du' e^{i(u-u'){k_y\over H_5}} e^{-i({\rm cosh} u
+{\rm cosh} u') mt} e^{-\epsilon |t|},
\label{eq:rateq}
\end{equation}
where we have inserted a Lorentz invariant convergence factor,
and include the volume factor needed to normalise the final states
(since $(2\pi)^3 {\delta}^3(0) =V)$.
The integrals are straightforwardly performed using the identity
\begin{equation}
\int_{-\infty}^\infty dx {{\rm cos}{k_y x} \over {\rm cosh}^2(x/2)} = 
{4 k_y \pi \over {\rm sinh} (k_y \pi/H_5)},
\end{equation}
to give
\begin{equation}
\langle{k_y},\vec{k}, {k'_y},\vec{k'},{\rm out}|0,{\rm in} \rangle = 
{i \mu^2 \over 2 m^2 H_5 V} 
(2\pi)^3 \delta^3(\vec{k}+\vec{k}') \delta_{k_y+k_y',0} \,
{(k_y/H_5) \over m^2 {\rm sinh}(k_y \pi /H_5)}.
\end{equation}
The probability for a transition from the vacuum to two particles
with momenta $k_y=\pm 2 \pi n/L$ and $\vec{k}$ within $d^3\vec{k}$ is therefore
\begin{equation}
{\mu^4\over 16} V {d^3 \vec{k}\over (2\pi)^2} {1\over |\vec{k}|^4}
{(k_y/H_5)^2\over {\rm sinh}^2(k_y\pi /H_5)}.
\label{pairvol}
\end{equation}
Dividing by the volume $V$ one obtains the
probability per unit volume
for creating such particle pairs. 
At fixed external momentum, the final density of pairs
is finite, as claimed. Furthermore, the $k_y\neq 0$ modes which
naively might be thought to be the most dangerous, are strongly
suppressed. As the rate $H_5$ of contraction of the extra dimension
is decreased, the production of these Kaluza Klein modes
becomes exponentially small, showing the existence of an
adiabatic limit.  The $k_y=0$ modes do not however display such 
a limit, and in fact the result for particle creation for these modes
is completely independent of $H_5$. 
We shall discuss how this behaviour is likely to be 
altered in string theory, below. 
 
The integration over $\vec{k}$ in (\ref{pairvol}) 
is infrared divergent. This is however an artefact of
the fact that the interaction term we introduced 
diverges as $t \rightarrow
\pm \infty$. 
In the situations of interest for the cyclic and ekpyrotic models, 
the extra dimension tends to a maximum size $t <<0$ and $t>>0$
and this would introduce an infrared cutoff in
$|\vec{k}|$ of order $T_C^{-1}$ where $T_C$ is the characteristic
time scale over which Milne-like behaviour holds.
The total number density of created 
particles with this simple $\varphi^2$ interaction is
ultraviolet finite. But the total energy density is
logarithmically divergent. This disease may be cured
by introducing the dilaton $\phi$ into the
nonlinear field interactions, which generically occurs in 
string theory. We have $ e^{\phi \sqrt{(d-2)/(d-1)}}\propto
|t|$, and each extra power of $|t|$ in the interaction
introduces an extra negative power of $|\vec{k}|$ in
the matrix element, or $|\vec{k}|^{-2}$ in the probability.
Conversely, if we introduce 
higher powers of $\varphi$ in the interaction, e.g. $\varphi^3$, 
this would boost the rate in the
ultraviolet, just because more particles are created in each process
and the phase space integral would involve a higher overall power of
$\vec{k}$. (Recall, there is no conservation of energy here since
we have explicitly broken time translation invariance).
Again, we can make the produced number or energy density
finite by introducing sufficient powers of the dilaton
coupling. However, as we shall now explain, we believe there should
be additional effects suppressing the rate of production of
particles with high momentum in string theory.

The origin of the power law fall-off in these calculations
may be traced to the
sharpness in time $t$
of the simple field theory interaction term we have introduced,
and the fact that this interaction cannot be correct
for very high momenta of the external particles.
The  Fourier transform of a function with a
sharp kink falls off only as a power law, so with this field theory 
interaction, 
high energy energy-nonconserving 
processes are only suppressed as a power law of the energy.
If interactions can be consistently introduced in string or M theory
on ${\cal M}_C$, we believe they will not show this sharp behaviour. 
Strings or membranes 
are never localised below a minimal length scale $l_s$, and 
one would expect
them not to see the extra dimension shrink below a minimal length scale
$l_s$. One could model this by replacing the factor 
$H_5 |t|L $, which is the length of the extra dimension, 
with $\sqrt{l_s^2 +H_5^2 L^2 t^2}$, which never
falls below $l_s$. 
If we make this replacement in the particle
production  
just computed, then the final particle density
is actually exponentially convergent in the
ultraviolet. Returning to (\ref{eq:rateq}), we see it is dominated by 
$u\sim u'\sim 0$. The effect of introducing the cutoff $l_s$ is
therefore roughly the replacement
\begin{equation}
\int dt |t| e^{-imt -\epsilon |t|} \rightarrow \int dt
 \sqrt{t^2+\delta^2} e^{-imt -\epsilon |t|},
\label{eq:cuto}
\end{equation}
where $\delta = l_s/(H_5 L)$. The left hand side 
equals $-2/m^2$ as $\epsilon \rightarrow 0$. This exhibits the power
law dependence of our answer for the amplitude 
above.
However, for large $m\delta$ the right hand side decays
exponentially in $m\delta$. To see this, compute the difference between the
left and right hand sides of (\ref{eq:cuto}), in which $\epsilon$ may be
set to zero from the outset. By integrating by parts, one can 
reduce the difference to 
\begin{equation}
-{2\over m^2} + \int_0^\infty {dt\over m^2} {\rm cos}(mt) { \delta^2
\over (t^2 +\delta^2)^{3\over2} }. 
\label{eq:cutoa}
\end{equation}
It follows that the integral equals minus the 
right hand side of (\ref{eq:cuto}).
The latter is 
a Hankel function of imaginary argument, which 
decays exponentially, as $e^{-m\delta} = e^{-m l_s/(H_5 L)}$, 
for large $m l_s$ (i.e. particle momenta well above the string scale),
or for small $H_5 L$ (i.e. a small contraction  speed of the 
extra dimension).

Two important things occur in this model.
First, 
the cutoff
is not at the string scale, it is at $|\vec{k}|\equiv m \sim (H_5 L)/l_s$.
More importantly, for fixed $|\vec{k}|$, the particle production 
becomes exponentially small as $H_5 L$ is lowered below 
$|\vec{k}| l_s$. 
This means that for 
$H_5L <<1$ and for modes of any 
fixed physical wavelength in the non-compact direction, 
there
is an adiabatic limit in which the extra
dimension can disappear and reappear with vanishingly small
particle production. It remains to be seen whether these two 
desirable features will 
survive in a complete string theoretic calculation.

\section{The d Dimensional Perspective}

The cyclic and ekpyrotic
universe scenarios represent attempts at consistent cosmologies 
based on M and string theory. The resolution of what happens at big
bang/big crunch singularities must be found within those theories.
If, for example, the extra dimension involved is the eleventh dimension
of M theory, as represented in the model of
Horava-Witten, then when the two boundary 
branes approach the theory reduces to weakly coupled heterotic
string theory. The full dynamics of the eleven dimensional theory
involves eleven dimensional supergravity and the associated super-membrane. 
Nevertheless, one hopes that at least for slow motions, the system
evolves quasi-adiabatically and at small brane separations one is
in the regime of ten dimensional string theory. Furthermore, since 
the string coupling constant vanishes as the extra dimension disappears, 
string interactions should be suppressed at the singularity itself.
However, as noted in Ref.~\ref{Seiberg}, the description of the bounce
in string theory may not be straightforward, because 
the ten dimensional string frame metric vanishes
at the boundary brane collision. 

In this section we shall attempt to describe not string theory
but quantum field theory from a purely $d$ dimensional perspective.
As in the string theory case, the $d$ dimensional metric will 
vanish at the singularity. Nevertheless we shall see that the 
singularity
may be traversed in a reasonably natural manner. 

The $d$ dimensional Einstein frame geometry corresponding to the 
$d+1$ Milne geometry
considered earlier is given by  
\begin{equation}
	ds^2=|H_5t|^{2/(d-2)}(-dt^2+{d\vec{x}}^2)
\end{equation}
Note that the proper time $t$ of the $d+1$ geometry is identical to the
conformal time $t$ of the $d$ dimensional Einstein frame geometry.
The equation of motion of a massless scalar field\footnote{Near $t=0$
massive particles behave like massless ones so it is sufficient to
consider the massless case.} is
\begin{equation}
	t^2 \ddot{\varphi}+ t \dot{\varphi}+t^2 \vec{k}^2 \varphi=0,
\end{equation}
just the equation for the $k_y=0$ mode of a $d+1$ dimensional
massless scalar. Using the higher dimensional point of view as our guide we are lead to believe
that the natural `in' and `out' states are
\begin{equation}
	\psi^{out,+}(t,x)=\psi^{in,+}(t,x)=\sqrt{\frac{\pi}{4H_5}} H^{(2)}_{0}(|\vec{k} t|)e^{i \vec{k}\cdot \vec{x}},
\end{equation}
and the corresponding Wightman function is
\begin{equation}
	G^+(x,x')= \int
	\frac{d^d\vec{k}}{(2 \pi)^d} \frac{\pi}{4H_5} H_{0}^{(2)}(|\vec{k}|(t-i\epsilon))
	 H_{0}^{(1)}(|\vec{k}| (t+i\epsilon)) e^{i\vec{k}\cdot(\vec{x}-\vec{x}')}
\end{equation}
How can we understand why this is natural purely from a $d$ dimensional
point of view? As already discussed in Section III, 
regulating the spacetime in the
sense of $|t| \rightarrow \sqrt{t^2 + \delta^2}$ will generically produce a
particle production that diverges as $\ln (H_5 \delta)$.
The fact that this divergence is identical for
each $\vec{k}$ suggests that it may be removed by a counterterm which is
local in $d$ dimensions.

\subsection{What renormalization?}

How should we remove the logarithmic divergence in the 
scalar field as it approaches the singularity? From the $d$ dimensional
perspective, we would like to follow the traditional renormalization program:
regularizing the theory, adding counterterms and finally
removing the regulator. But we need to discuss what 
form the renormalization should take. 

As mentioned above, in the limit when $t$ tends to zero, the dynamics
of the field are dominated by the kinetic term. This term
possesses a symmetry $t \rightarrow \lambda t$ (which is just a 
translation in conformal time).
The scalar
field tends to $\varphi \sim A +B {\rm ln}|H_5 t|$, and its momentum $\pi$
tends to $H_5 B$.  Rescaling time as above has the effect of
increasing $\varphi$ by $B {\rm ln} \lambda$, or 
$ H_5^{-1} \pi{\rm ln} \lambda$.
Therefore it is very natural
to seek to exploit such a shift in order to remove the divergence in 
$\varphi$. Since $\pi$ is asymptotically a constant, 
one can simply match it 
across $t=0$. But to make $\varphi$ finite, we need to 
redefine it via
$\varphi \rightarrow \varphi+
C \pi$, across the $t=0$ surface in the regulated theory, with
the 
constant $C$ chosen 
to obtain finite correlators for $\varphi$ and $\pi$ 
for $t>0$, as the regulator is removed. 

The  shift $\varphi \rightarrow \varphi+
C \pi$, $\pi \rightarrow \pi$, 
is a canonical transformation. It can be
implemented by adding a 
a local counterterm to the Hamiltonian, which acts at $t=0$
to produce an additional unitary transformation 
taking the incoming to the outgoing quantum state.
We shall show below that demanding the vacuum be both Hadamard and 
time reversal invariant uniquely fixes the value of the constant
$C$.

\subsection{Dimensional Regularization}

Before we can renormalize the $d$-dimensional theory we must first
regularize it. We shall use a regularization which makes 
both the field and its canonical momentum finite at $t=0$, 
but which allows the background to remain a solution of
the field equations everywhere except at $t=0$. 

The dimensional regularization we use
relies upon 
the generalization of the $1+1$ dimensional Milne universe 
we have so far studied to a $1+n$ dimensional Milne universe.
Like their $1+1$ dimensional cousin these are just a re-writing
of Minkowski spacetime, but in dimensions $n>1$ their 
constant time slices are negatively curved 
hyperboloids $H_n$. We therefore consider the following 
vacuum Einstein's equations in $d+1$ dimensions,
\begin{equation}
	ds^2=-dt^2+H_5^2 t^2 dH_n^2  +\sum_{i=1}^{d-n} dx_i^2,
\label{eq:lineel}
\end{equation}
where $d H_n^2$ is the line element on 
$H_n$. This is a solution of the $d+1$ dimensional field 
equations if the curvature (Ricci) scalar of the 
$H_n$ is
 $-n(n-1)H_5^2/2$.
From the $d-n$-dimensional point of
view, we are considering fields which are constant on the $H_n$, 
we have Einstein gravity plus a scalar field $\phi$ representing
the `scale factor' of the $H_n$.\footnote{ Parenthetically,
we remark that even though 
$H_n$ is noncompact, there is a natural splitting between 
the homogeneous modes and the non-constant modes, because the
Laplacian on $H_n$ has a smallest nonzero eigenvalue equal to 
the space curvature. Therefore provided we are considering $d-n$ dimensional
energy
scales much smaller than $H_5$, we may consistently neglect the non-constant
modes.} The spatial curvature term in the $d+1$ dimensional Einstein
action leads to a potential for $\phi$ in the 
dimensionally reduced $d-n$ dimensional action,
proportional
to $n(n-1)$, and positive for $n>1$. 
The idea of the regularization we use is to analytically
continue this reduced theory in $n$, and 
ultimately take the limit as $n\rightarrow 1$.

The action for
a scalar field which is homogeneous on the $H_n$ is just
\begin{equation}
\label{eq:regaction}
-{1\over 2} \int dt |H_5 t|^n \eta^{\mu \nu} \partial_{\mu}\varphi
\partial_\nu \varphi.
\end{equation}
 For $n=1-\epsilon$, with $\epsilon $
small and positive, it turns out that 
the field $\varphi$ and its canonical momentum
$\pi$ are both finite at $t=0$ and therefore both can be
simultaneously matched across $t=0$.
To see this, note that as
$t \rightarrow 0$ the scalar field equation is approximated
by 
\begin{equation}
	t^2 \ddot{\varphi}+(1-\epsilon) t \dot{\varphi}\approx 0,
\end{equation}
with general solution $\varphi\sim A+B (H_5 t)^{\epsilon}$. The momentum
conjugate to $\varphi$ is given by $\pi = |H_5 t|^{1-\epsilon} \dot{\varphi} \sim 
\epsilon H_5 B$, constant in this limit. 
The remarkable feature is that for positive 
$\epsilon$,  $\varphi$ is also finite at
$t=0$, enabling us to match across $t=0$.
(The limiting case $\epsilon=0$, which we studied 
before, 
can be obtained by expanding $(H_5 t)^{\epsilon} \approx
 1+\epsilon \ln(H_5 t)$ for small $\epsilon$,
and redefining the constants.) 

Whilst it is important that we have constructed the regularized
backgrounds as solutions of the field equations everywhere
except at $t=0$, for the purposes of this section all that really
matters is that we have regularized the action for the scalar
$\varphi$ to (\ref{eq:regaction}) for $t<0$ and for $t>0$
so that we can match $\varphi$ and $\pi$ across $t=0$, with
the introduction of local counterterms at that point.

\subsection{Matching Modes across the Singularity}

The regularized field equation for the $y$-independent modes is 
\begin{equation}
	t^2 \ddot{\varphi}+(1-\epsilon) t \dot{\varphi}+m^2 t^2\varphi=0,
\end{equation}
with $m^2\equiv \vec{k}^2$ as before. The 
general solution is $\varphi = (mt)^{(\epsilon/ 2)} \chi(mt)$,
with $\chi$ a Bessel function of order $\epsilon/2$.

As $t$ tends to zero, solutions to the regularized equation tend to 
the form $\varphi\sim A+B|\vec{k} t|^{\epsilon}$, with
$A$ and $B$ constants. Thus as claimed above, both $\varphi$ and
$\pi \equiv |H_5 t|^{1-\epsilon} \dot{\varphi}$ tend to constants as $t$ tends to
zero. Let
$(A^-,B^-)$ and $(A^+,B^+)$ denote the values of these constants for
$t <0$ and $t >0$. Matching $\varphi$ and $\pi$ at $t=0$ gives $A^+=A^-$ and
$B^+=-B^-$. Thus the asymptotic solutions for
small $|\vec{k} t|$ are
\begin{equation}
	\varphi\sim A+B {\rm sign} (t) |\vec{k} t|^{\epsilon}.
\end{equation} 
Using the asymptotic form of
the Hankel functions for small argument and redefining the constants
we find that the general solution for all $t$ is
\begin{eqnarray}
&&	\varphi=\cr
&& |\vec{k} t |^{\epsilon/2} [C\left(e^{i\epsilon \pi/2} H^{(1)}_{\epsilon/2}(|\vec{k} t|)+ e^{-i\epsilon
\pi/2}H^{(2)}_{\epsilon/2}(|\vec{k} t|)\right) +
	D {\rm sign} (t)\left(H^{(1)}_{\epsilon/2}(|\vec{k} t|)+
	H^{(2)}_{\epsilon/2}(|\vec{k} t|)\right)].
\end{eqnarray}
As before defining the in and out states to be $\varphi_{in}=A_{\epsilon/2}
H^{(1)}_{\epsilon/2}(-|\vec{k}| t)$ for $t <0$ and $\varphi_{out}=A_{\epsilon/2}
H^{(2)}_{\epsilon/2}(|\vec{k}| t)$ for $t >0$ we find the following relation
between the in and out states:
\begin{equation}
\varphi_{in}=\alpha \varphi_{out}+\beta \varphi^*_{out}, \qquad 
	\alpha =-i {\rm cosec}(\pi \epsilon/2) e^{-i\pi \epsilon/2}, \qquad
	 \beta=-i \cot(\pi \epsilon/2).
\end{equation}
Let us consider the limit $\epsilon \rightarrow 0$. In this limit
$\alpha \rightarrow -2i/(\pi \epsilon) $ and $\beta \rightarrow -2i/(\pi
\epsilon)$. This implies infinite particle production, as
we obtained before with the simple cutoff. In fact the analogy is 
\begin{equation}
	\frac{1}{\epsilon} \approx \ln(\frac{1}{H_5 \delta}).
\end{equation}
This relationship is similar to that obtained in ordinary
renormalization where $1/(d-4)$ divergences in dimensional regularization
corresponds to the $\ln(\Lambda)$ divergences obtained with 
a UV cutoff.

\subsection{Regularization Independence}

In the previous section we have seen that the logarithmic 
divergences of the naive cutoff regularization used
earlier manifest themselves as $1/\epsilon$ divergences in the dimensional
regularization scheme. We now want to 
renormalize these divergences, by absorbing the divergent
pieces in local counterterms added to the action.

In dimensional regularization the minimal subtraction scheme is simply
to throw away $1/\epsilon$ divergences. Here, one must be more careful
since there is a danger of violating unitarity. 
We are able to
perform the subtraction via a unitary transformation as follows.
First write the Bogoliubov transformation in the form
\begin{equation}
\left(\begin{array}{cc} \varphi_{out} \\ \varphi_{out}^* \\ \end{array}
\right)
=\left(\begin{array}{cc} \alpha & \beta \\ \beta^* & \alpha^* \\ \end{array} \right)
\left(\begin{array}{cc} \varphi_{in} \\ \varphi_{in}^* \\ \end{array}
\right)
=\left(\begin{array}{cc} -i{\rm cosec}(\pi
	\epsilon/2)e^{-i\pi
	\epsilon/2} & -i\cot(\pi
	\epsilon/2) \\  i\cot(\pi
	\epsilon/2)& i {\rm cosec}(\pi
	\epsilon/2)e^{i\pi
	\epsilon/2} \\ \end{array} \right)
\left(\begin{array}{cc} \varphi_{in} \\ \varphi_{in}^* \\ \end{array}
\right),
\end{equation}
Now redefine the out state by means of the following Bogoliubov
transformation
\begin{equation}
\left(\begin{array}{cc} {\varphi}'_{out} \\ {{\varphi}'_{out}}^* \\ \end{array}
\right)
=\left(\begin{array}{cc} 1+\frac{2i}{\pi \epsilon} & \frac{2i}{\pi
\epsilon} \\ -\frac{2i}{\pi \epsilon} &  1-\frac{2i}{\pi \epsilon} \\ \end{array} \right)
\left(\begin{array}{cc} \varphi_{out} \\ \varphi_{out}^* \\ \end{array}
\right),
\end{equation}
Then in the limit $\epsilon \rightarrow 0$ we find
\begin{equation}
\left(\begin{array}{cc} {\varphi}'_{out} \\ {{\varphi}'}_{out}^* \\ \end{array}
\right)
=\left(\begin{array}{cc} -1 & 0 \\ 0 & -1 \\ \end{array} \right)
\left(\begin{array}{cc} \varphi_{in} \\ \varphi_{in}^* \\ \end{array}
\right),
\end{equation}
but since the $1/\epsilon$ divergences have been removed by means of a
Bogoliubov (unitary) transformation, unitarity is clearly
preserved. 

The overall minus sign aquired by the mode functions 
${\varphi}'_{out}$ after passing through $t=0$ is an
unobservable phase in the quantum mechanical wavefunction. However,
it is interesting to compare what we have done here to what was done
in the earlier methods of analytic continuation. Here we
have insisted that $\pi$ be continuous, and have found that,
after renormalization, $\varphi$
must change sign. Whereas there, the continuation produced
a divergent part of the field which was even
with a corresponding momentum $\pi$ which was odd, and had a jump at
$t=0$. The two methods
differ by an overall minus sign but this is just a phase and
is physically irrelevant.

In fact, we can easily compute the magnitude of the jump in
$\varphi$ in the regulated theory, using the small argument form 
for the Hankel functions. We find the result that after 
the above renormalization, 
\begin{equation}
\varphi(0^+)-\varphi(0^-) = -{2 \over \epsilon H_5} \pi(0). 
\label{eq:jump}
\end{equation}
But this is precisely a canonical
transformation corresponding to the shift symmetry of the massless
homogeneous scalar field described earlier. This justifies the
conjecture than the divergent piece can be removed by means of a
canonical tranformation. 
The result is that this subtraction scheme predicts no particle
creation for a massless scalar field, exactly the result of
the Minkowski embedding matching. So we have reached the same
conclusion from the $d$-dimensional persepective. 

\subsection{Unitary Evolution and Regulating the Hamiltonian}

The previous result is in a sense a trivial one. We can always
redefine the in state by means of a Bogoliubov transformation such
that $|out>=|in>$. However, the critical point is that the 
Bogoliubov transformation we used was momentum independent and so corresponds
to a local unitary transformation.
Any canonical/Bogoliubov transformation can be thought of as an instantaneous
interaction term in the Hamiltonian. Let $\Delta S$ denote the `impulse'
(which has dimensions of action), acting at time $t$,
so the time dependent Hamiltonian is 
\begin{equation}
	H=H_0+\delta(t) \Delta S.
\end{equation}
Then the unitary evolution operator $U_H(t,t')$ for $t'<0$ and $t>0$ may be expressed
formally as
\begin{equation}
	U_H(t,t')=T\exp(-i\int_{t'}^t H(t) dt) 
	=U_H(t,0^+) \exp(-i\Delta S ) U_H(0^-,t'),
\end{equation}
in other words unitary evolution with Hamiltonian $H_0$ up until
time $t=0^-$, a unitary transformation $\exp(-i\Delta S)$ followed by unitary
evolution with $H_0$ again. In particular if $\Delta S $ is quadratic in the
fields and momenta then $\exp(-i\Delta S)$ performs a canonical
transformation of the form described in Appendix 1. Comparing
with the result (\ref{eq:jump}) above, we see that $\Delta S$ is
the local counterterm
\begin{equation}
\Delta S =-\frac{1}{\epsilon H_5} \int d^3 \vec{x} \pi^2(\vec{x},0).
\label{eq:deltas}
\end{equation}
In principle we have freedom to add other local counterterms, a
freedom leaving us with a 3 parameter family of matching
conditions. 
However, as we explain in Appendix 2. the above choice 
is the unique one that yields a 
Green function which is both of Hadamard form and time reversal invariant,
i.e. such that the 
Wightman function transforms under the time reversal operator 
$T$ by $G^+(Tx,Tx')=G^-(x,x')$.

Another way of seeing why  $\Delta S $ given in (\ref{eq:deltas})
is the correct counterterm is
to consider the bare Hamiltonian in the dimensional regulation scheme. Near
the singularity the leading contribution to the Hamiltonian is from
the kinetic term. Computing from the action, (\ref{eq:regaction}),
we find the kinetic term contributes 
 $H=\frac{1}{2} \int d^3\vec{x} 
\pi^2/|H_5 t|^{1-\epsilon}$. The unitary evolution operator 
depends on time ordered 
products of integrals of this Hamiltonian. 
Consider the contribution of the region in the vicinity of $t=0$, to 
$\int H dt$ for
$|t|<T$. Using the fact that $\pi$ is nearly constant, we
may approximate this for small $T$ as
\begin{equation}
{1\over 2}
\int_{-T}^{T} |H_5 t|^{-(1-\epsilon)} dt  \int d^3 \vec{x} \pi^2(\vec{x},0)
= {(H_5 T)^\epsilon \over \epsilon  H_5}
\int d^3 \vec{x} \pi^2(\vec{x},0).
\end{equation}
Now taking the limit as $\epsilon \rightarrow 0$, we obtain a 
leading term of precisely minus $\Delta S$ given in (\ref{eq:deltas}). 
Thus we see that the counterterm we have introduced is just the
minimal subtraction needed
to obtain a well defined
limit for the time evolution operator as $\epsilon$ is taken to zero.

Let us briefly discuss the issue of general covariance. In this
paper we are working in a fixed background and
not including gravitational effects. There is a preferred slicing of this
background and therefore it is allowable to have a counterterm which
is explicitly $t$-dependent. However, when gravity is included, 
there should be no preferred coordinate. How should we think
of the counterterm $\Delta S$ in that case? The point is that
from the $d$ dimensional point of view, there is a scalar field
which yields a coordinate invariant time-slicing of the geometry,
namely the dilaton $\phi$ which we have mostly neglected.
If we rewrite $\Delta S$ in terms of the dilaton field, it
is then coordinate invariant and can be used in a generally
covariant treatment.

\section{Quantum Fields in de Sitter Space }

In this section we want to discuss an interesting formal
correspondence between quantum fields on ${\cal M}_C$
and on de Sitter space. 
The spacetime ${\cal{M_C}} \times
R^{d-1}$ is conformal to the 
non-singular spacetime $S^1 \times dS^{d}$, as may immediately be
seen
from the metric
\begin{equation}
	ds^2=H_5^2 t^2 [dy^2 +\frac{1}{(H_5 t)^2}(-dt^2+d \vec{x}^2)],
\end{equation}
where the latter term is just the metric on de Sitter space in the
flat slicing. The parameter 
$H_5$ plays the role of the Hubble constant of the $d$
dimensional de Sitter space. Figure 5 shows the correspondence between
the two halves of the Milne universe, $L$ and $U$, and the
two flat-sliced regions of de Sitter space, one of which
covers the region containing past
timelike infinity, $t=0^+$ up to the co-ordinate horizon
at $t=+\infty$, and the other which covers the region 
from the horizon $t=-\infty$ to future
timelike infinity $t=0^-$. 
In the de Sitter geometry,
starting from a generic point in region $L$, 
it takes an infinite proper time to 
reach $t=0^-$, but 
in ${\cal M}$ only a finite proper time is needed.
The key point in this correspondence is that 
the two surfaces $t=0^-$ and $t=0^+$ meeting at
the singularity of ${\cal M}_C$ are
mapped to the two spacelike surfaces representing 
timelike past and future infinity in de Sitter space.

\begin{figure}
{\par\centering
\resizebox*{6.in}{2.9in}{\includegraphics{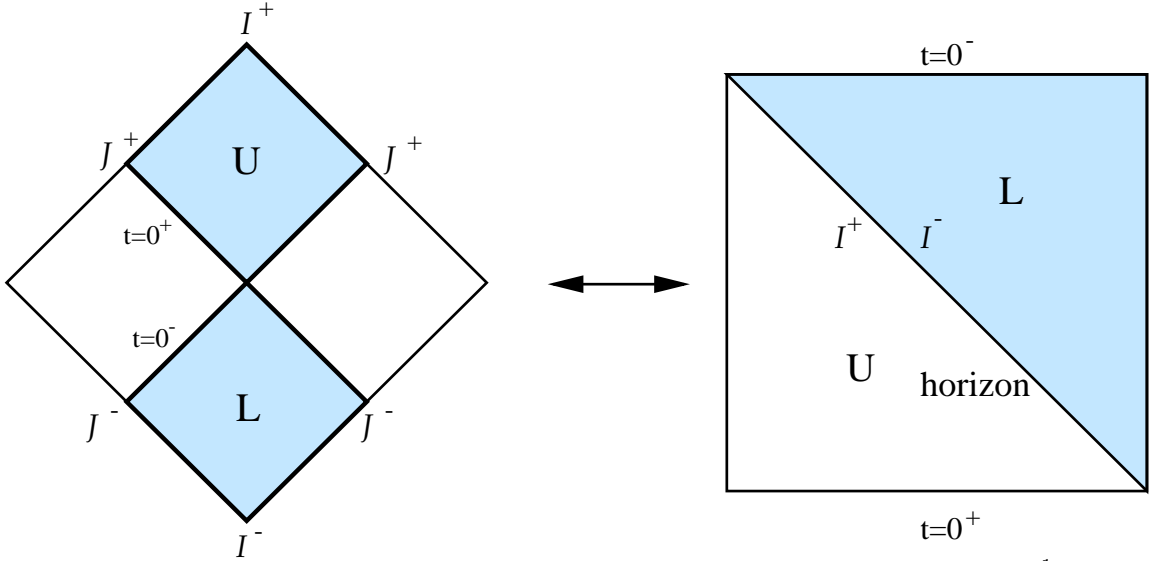}} \par}
\caption{The Milne spacetime is both locally flat (left diagram)
and conformally $S^1 \times$ de Sitter (right diagram, with the 5th
dimension suppressed). In the de Sitter picture
passing through the singularity corresponds to matching future
timelike infinity in de Sitter to past timelike infinity. }
\end{figure}

A peculiarity of this map is that the collapsing region $L$ of the Milne
geometry is mapped to the expanding 
region $L$ 
of de Sitter
space. Nevertheless the arrow of time always points in the direction
of increasing $t$.

De Sitter space can also be globally covered by the closed slicing
co-ordinates, so-called because the Cauchy surfaces are spheres
$S^{d-1}$. If $d^2\Omega_{S^{d-1}}$ denotes the metric on an $S^{d-1}$
with radius $H_5$ then the de Sitter metric is
\begin{equation}
	ds^2 =-d\tau^2+\cosh^2 (H_5\tau) d^2\Omega_{S^{d-1}}
\end{equation}
There is no co-ordinate singularity in these co-ordinates which 
provide an unambiguous method for matching fields from
 $\cal{I}^-$ to $\cal{I}^+$. There exists a
unitary operator $U(\tau_1,\tau_2)$ which generates time evolution
from $\tau_2$ to $\tau_1$. This operator satisfies the Schrodinger
equation
\begin{equation}
	i\frac{\partial}{\partial \tau_1}U(\tau_1,\tau_2)=H(\tau_1) U(\tau_1,\tau_2)
\end{equation}
where $H(\tau)$ is the time-dependent Hamiltonian. The 
$S$ matrix is defined as the unitary operator 
\begin{equation}
	S=\lim_{\tau_1 \rightarrow \infty, \tau_2 \rightarrow -\infty}
	U(\tau_1,\tau_2)=T e^{-i \int_{-\infty}^{\infty} H(\tau) d\tau},
\end{equation}
which in the Milne picture corresponds to an $S$ matrix
matching from $t=0^+$ to $t=0^-$ or equivalently,
\begin{equation}
	|\psi,t=0^+>=S^{\dagger} |\psi,t=0^->
\end{equation}
Free field theory on de Sitter spacetime therefore provides us with
yet another matching prescription for Milne spacetime.

Since the spacetime ${\cal{M_C}} \times R^{d-1}$ is locally flat then a
massless minimally coupled scalar field is identical to a massless
conformally coupled scalar field. In $d+1$ dimensions the conformal
coupling term in the Lagrangian is $-\frac{(d-1)}{8d} R \varphi^2$. 
After a conformal transformation we obtain a massless conformally
coupled scalar field on $S^1 \times dS^{d}$. The Ricci scalar on $S^1 \times
dS^{d}$ is simply a constant $R=d(d-1) H_5^2$ which means we may
reinterpret the scalar field as a minimally coupled scalar field with
mass $m^2=(d-1)^2 H_5^2/4$. On Kaluza-Klein compactification over
the $S^1$ we obtain a tower of scalar fields on $dS^d$ with masses
$m^2=(d-1)^2 H_5^2/4+(2\pi n/L)^2$.

As we explained earlier the KK zero mode of a minimally coupled scalar
on ${\cal{M_C}} \times R^{d-1}$ is identical to a minimally coupled
scalar on the lower dimensional Einstein frame geometry. This in turn
is equivalent to a scalar field of mass $m^2=(d-1)^2 H_5^2/4$ on
$dS^d$. This result would seem surprising had we not used the higher
dimensional geometry since although the $d$ dimensional geometry is
conformal to $dS^d$, a minimally coupled scalar field is not
conformally invariant.

In de Sitter space there is a one-parameter family of de Sitter
invariant vacua\cite{ballen}.
 A unique choice, known as the Bunch-Davies vacuum\cite{Birrell},
is obtained if we also demand that the Feynman propagator be of the
Hadamard form (see Appendix 2). 
A de Sitter invariant measure of the distance between
two points is provided by the variable $z=1+((t-t')^2-(\vec x -\vec
x')^2)/2tt'$. Although we have expressed it in terms of flat slicing
co-ordinates, $z$ is globally defined and has the same definition
regardless of which patch $t$ and $t'$ are in. This means that the
Feynman propagator in the Bunch-Davies vacuum can be defined globally
in terms of the hypergeometric function\cite{Birrell}
\begin{equation}
	G_F=H_5^2\frac{\Gamma((d-1)/2+\nu)\Gamma((d-1)/2-\nu)}{8 \pi^2} {}_2F_1[(d-1)/2-\nu,(d-1)/2+\nu;2,\frac{1}{2}(1+z-i\epsilon)]
\end{equation}
where $\nu$ depends on the Kaluza-Klein mode number $n$, 
$\nu = 2\pi in/H_5 L$. 

We are interested in calculating the $S$ matrix to determine the
matching condition implied by the above propagator. At the free field
level it is sufficient to consider how a single particle state
evolves. The Feynman propagator allows us to evolve the
positive frequency part of a scalar field
from one Cauchy surface $t'=$ constant, to any point in its causal
future,
\begin{equation}
\varphi^+(x,t)=i\int d^{d-1} 
\vec{x}' (Ht')^{-d}G_F(x,t;x',t') 
(\overrightarrow{ \partial_{t'}} -\overleftarrow{\partial_{t'}})
 \varphi^+(x',t')
\end{equation}
A simple calculation shows that a positive frequency WKB in state in
the region $t \rightarrow +\infty$ evolves to a positive frequency WKB
state in the region $t \rightarrow -\infty$. The reason for this can
be seen immediately from the conformal diagram since $\cal{I^+}$ and
$\cal{I^-}$ are identified as the same coordinate horizon.

This result can be understood more clearly by realizing that in
analogy with Minkowski, we can define particles in de Sitter as
representations of the de Sitter group. Since these representations
are globally defined and  we choose a vacuum that respects the de
Sitter symmetry, then it is clear that in the absence of interactions
there can be no particle creation in de Sitter spacetime.

At this point we should make clear that this is not in contradiction
with the usual statement that there is a thermal distribution of
particles in de Sitter space. This description arises because of a
different definition of particles, one which is more appropriate to
the static patch surrounding an observer's world-line. The definition
of particles in a curved spacetime is observer dependent, but
the evolution of fields is observer-independent The
Bunch-Davies vacuum is the vacuum in which globally there is no
particle creation according to the representation theory definition, but
where locally an observer sees a thermal bath of particles.

In conclusion we have reached the same results as before for free field theory,
using the de Sitter
picture. When one includes interactions, a careful track of the
non-minimal couplings must be taken into account. For instance
$\lambda \varphi^4$ theory in the 4d Einstein frame geometry will
correspond to massive $\lambda'(t) \varphi^4$ theory on de Sitter where
$\lambda'(t)$ is a new time-dependent coupling constant. Ultimately,
it is not
clear to us that the Milne-de Sitter correspondence 
be a useful guide, because the problem of string theory
on de Sitter space is probably a harder problem than that of
string theory on the
Milne spacetime.

\section{Conclusion}

In this paper we have shown that it is possible to define
free quantum fields on the compactified Milne universe in a
consistent and 
unambiguous 
manner. We have made limited progress in studying interactions, 
and how these lead to particle production.
The density of particles produced at fixed external momentum is
finite at tree level. 
The integrated density was also found to be finite 
provided the dilaton dependence of couplings caused
them to vanish sufficiently rapidly with $t$. We suggested how
an adiabatic limit, in which particle 
production would be exponentially small for
small $H_5L$,  might emerge in string theory.
We also pointed out 
connections with quantum field theory on de Sitter spacetime, 
which may well be interesting in their own right.

Certainly, much remains to be done to explore quantum fields
on the Milne and compactified Milne universe. The 
methods used here could be extended to include
gravitational backreaction, at least for linearised gravity, 
to follow cosmological perturbations through the singularity.
We shall 
report on a study of loop diagrams for scalar field interactions
on ${\cal M}_C$
in the near future.
A major challenge remaining is to extend these ideas
to string theory and M theory. 

\bigskip

\noindent
{\bf Acknowledgements:} 
\smallskip
We thank Martin Bucher, Ruth Durrer, 
Steven Gratton, David Gross, James Hartle, Stephen Hawking,
Joe Polchinski, Fernando Quevedo, Nati Seiberg, 
Paul Steinhardt, Gabriele Veneziano 
and  Toby Wiseman for many useful discussions.
AJT
acknowledges the support of an EPSRC studentship. The work
of NT is supported by PPARC (UK).

\bigskip
{\bf Note Added:}
Since this article was submitted to the archive a number of related
articles condisidering fields/strings on backgrounds with cosmological
singularities have appeared\cite{recent}.

\vfill\eject
\section{Appendix 1: The Relationship between Canonical and Bogoliubov
Transformations}

The group of Bogoliubov transformations is identical to the group of
linear canonical transformations. For a 2 dimensional phase space
these form the group $Sp(2)$, ie those $2 \times 2$ matrices that
satisfy
\begin{equation}
	A^T \Omega A = \Omega,
\end{equation}
where $\Omega= \left(
        \begin{array}{cc}
               0 & -1 \\
               1 & 0 \\
        \end{array}
       \right)$.
$Sp(2)$ is a real form of $SU(2)$ and consequently we can write any
        generator $A$ in terms of the Pauli matrices
        $\sigma_{\pm}=\frac{1}{2}(\sigma_1 \pm \sigma_2)$, $\sigma_3$.
\begin{equation}
	A=A_+(a)A_-(b)A_3(c)=e^{a \sigma_+}e^{b \sigma_-}e^{c \sigma_3}
\end{equation}
If $\left(\begin{array}{cc} q \\ p \\ \end{array} \right)$ denotes an
arbitrary vector in phase space then the transformed vector
$\left(\begin{array}{cc} q' \\ p' \\ \end{array} \right) =A
\left(\begin{array}{cc} q \\ p \\ \end{array} \right)$ is given by
\begin{eqnarray}
	\nonumber & &
A_+ \left(\begin{array}{cc} q \\ p \\ \end{array} \right) =
\left(\begin{array}{cc} q+ap \\ p \\ \end{array} \right)\\ \nonumber & & 	
A_- \left(\begin{array}{cc} q \\ p \\ \end{array} \right) = \left(\begin{array}{cc} q \\ p+bq \\ \end{array} \right)\\ & &
A_3 \left(\begin{array}{cc} q \\ p \\ \end{array} \right)=
\left(\begin{array}{cc} e^c q \\ e^{-c} p \\ \end{array} \right).
\end{eqnarray}
In quantum mechanics $q$ and $p$ are replaced by operators $Q$ and $P$
satisfying the Heisenberg algebra
\begin{equation}
	[Q,Q]=[P,P]=0 , \qquad [Q,P]=i.
\end{equation}
The canonical transformations $A_+$, $A_-$ and $A_3$ can be
represented by 3 unitary transformations 
$U_+$, $U_-$ and $U_3$ given by
\begin{eqnarray}
	\nonumber & &
	U_+(a)=\exp(\frac{1}{2} ia P^2), \\ \nonumber & &
	U_-(b)=\exp(-\frac{1}{2} ib Q^2), \\ \nonumber & &
	U_3(c)=\exp(\frac{1}{2} ic (PQ+QP)). \\ & &
\end{eqnarray}
These follow as a simple consequence of the Heisenberg algebra. We can
define creation and annihilation operators in the usual way
$a=\frac{1}{\sqrt{2}}(Q+iP)$, $a^{\dagger}=\frac{1}{\sqrt{2}}(Q-iP)$
and consequently an arbitrary canonical transformation corresponds to
an arbitrary redefinition of $a$ and $a^{\dagger}$, ie. a Bogoliubov
transformation.

In field theory an operator valued field can be expressed in terms of
creation and annihilation operators as
\begin{equation}
	\varphi(x)= \sum_i a_i \psi_i(x) +a_i^{\dagger} \psi^*_i (x)
	= \sqrt{2}  \sum_i Q_i Re(\psi_i(x)) -P_i Im(\psi_i (x))
\end{equation}
with $\psi_i(x)$ and $\psi^*_i(x)$ the positive and negative frequency
mode functions, normalised according to the Klein-Gordon inner product.
It is usual to write a Bogoliubov transformation as a transformation acting
on these modes: $\psi'_i(x)=\alpha \psi_i(x)+\beta \psi^*_i(x)$,
which preserves the Klein-Gordon norm if $|\alpha|^2-|\beta|^2=1.$
Since the field $\varphi(x)$ is invariant under this transformation,
the creation and annihilation operators must transform under the inverse
Bogoliubov transformation,
\begin{eqnarray} \nonumber
	a'_i&=&\alpha^*a_i-\beta^*  a^{\dagger}_i \\ 
	a'^{\dagger}_i&=&\alpha a^{\dagger}_i-\beta  a_i.
\end{eqnarray}
If we re-write this in terms of coordinates and momenta, then in the
above notation, the Boguliubov transformation  
corresponds to the linear canonical transformation on the operators $(Q_i,P_i)$
\begin{equation}
	A= \left(
        \begin{array}{cc}
               Re(\alpha-\beta) & Im(\alpha+\beta) \\
               Im(\beta-\alpha) & Re(\alpha+\beta) \\
        \end{array}
       \right)
\end{equation}
It is simple to check that $det A=|\alpha|^2-|\beta|^2 =1$ as
required. This formula gives the precise map between canonical
transformations and a Bogoliubov transformations.

\section{Appendix 2: Hadamard form of the propagator}

In the description of quantum fields on a curved spacetime, a natural
and common restriction on the choice of vacuum is to impose that the
Feynman propagator should be of Hadamard form. More precisely this
means that in the coincidence limit $x \rightarrow x'$, $G_F(x,x')$
has the same singularity structure as the Feynman propagator on flat
space. A physical motivation for this choice of vacuum is that two
observers located at nearby points 
$x$ and $x'$ should not be able to tell if they are on a
curved space or Minkowski space by information sent between them.
Since we are
considering the coincidence limit $x \rightarrow x'$ the distinction
between the various types of Green's functions is not relevant and it
is common to work exclusively with the Hadamard function defined by
\begin{equation}
	G_H(x,x')=G^+(x,x')+G^-(x,x')=-2 {\rm Im} G_F(x,x'), 
\end{equation}
where 
$G^+(x,x')$ denotes the Wightman function $<0|\varphi(x)\varphi(x')|0>$ and
$G^-(x,x')$ is its complex conjugate.
Now suppose we are interested in the Green's functions on a spacetime
which is invariant under the discrete symmetries of time reversal $T$
and parity $P$. This is true of all the spacetimes we have considered
which can be seen by a simple inspection of their metrics. In
particular then the spacetimes are invariant under the combined
symmetries $PT$. This operator is anti-unitary and modes $\psi_i(x)$
may be decomposed into its eigenstates, which with an appropriate 
choice of phase can be chosen to obey
\begin{equation}
	\psi(PT x)= \psi^*(x).
\end{equation}
Given one such eigenstate, let us construct an 
arbitrary momentum-independent
 Bogoliubov transformation, to the mode function
\begin{equation}
	\psi'(x)=\alpha \psi(x)+\beta \psi^*(x).
\end{equation}
Then we find that
\begin{equation}
	\psi'(PT x)= \alpha \psi^*(x) +\beta  \psi(x).
\end{equation}
If we demand that $\psi'(x)$ also be an eigenstate of $PT$ with
eigenvalue $\eta$ (which can be an arbitrary complex phase), 
then we must have
\begin{equation}
	\alpha^* = \eta \alpha , \beta^* = \eta \beta,
\end{equation}
relations which can only be satisfied if 
Im$(\alpha \beta^*)$
is zero, a result we shall use in a moment. 

Given a particular set of positive frequency modes $\psi_i(x)$ which
are time reversal invariant $\psi_i(PTx)= \psi_i(x)^*$ and for which the
vacuum is Hadamard we can construct the Hadamard function as,
\begin{equation}
	G_H(x,x')=\sum_i \psi_i(x) \psi^*_i(x')+ c.c.
\end{equation}
Now define a new vacuum by means of a constant Bogoliubov
transformation of the first vacuum
\begin{equation}
	\psi'_i(x)=\alpha \psi_i(x)+\beta \psi_i^*(x).
\end{equation}
The Hadamard function in the new vacuum is given by
\begin{eqnarray}
	\nonumber & &
	{G}'_H(x,x')=(|\alpha|^2+|\beta|^2)G_H(x,x') 
	+2 Re(\alpha \beta^*) G_H(x,PTx') -2 Im(\alpha \beta^*) \Delta(x,PTx')
\end{eqnarray}
The commutator function 
\begin{equation}
	[\varphi(x),\varphi(x')]=G^+(x,x')-G^-(x,x')=i\Delta(x,x')
\end{equation}
does not contribute to the
Hadamard singularity structure since it is vacuum independent. The
singularity of $\Delta(x,x')$ occurs for null separated points
only. However, the
commutator ${G}'_H(x,x')$ in the new vacuum will have singular
behaviour in the coincidence limit $x' \rightarrow PTx$. But one of
the requirements of the Hadamard vacuum is that the propagator only has
a singularity as $x \rightarrow x'$. Consequently if we demand that the new
vacuum is Hadamard then we must have Re$(\alpha \beta^*)=0$.

The conditions derived in the previous two paragraphs together
imply that $\beta=0$. In other words, the
requirement of
$PT$ invariance and Hadamard form uniquely picks out the vacuum. In
fact all the vacua we have considered in this paper are trivially $P$
invariant, $G^+(Px,Px')=G^+(x,x')$, and so we only
need to additionally impose the restriction of $T$ invariance.

To put this in the more familiar setting of quantum fields on de Sitter
space, the condition that the vacuum
propagator be de Sitter invariant automatically picks out
a vacuum invariant under $PT$ since $PT$ is a discrete subgroup of the
full de Sitter symmetry. Then, as is well known, the additional
requirement of Hadamard form uniquely picks the vacuum as the standard
Euclidean, or 
Bunch-Davies vacuum.

\end{document}